# Fermi surface of magnetic kagome compound GdV$_6$Sn$_6$ investigated using de Haas van Alphen Oscillations

C. Dhital[1*], G. Pokharel[2], B. Wilson[1], I. Kendrick[1], M.M. Asmar[1], D. Graf[3], J. Guerrero-Sanchez[4], R. Gonzalez-Hernandez[5], and S.D. Wilson[2]

[1] *Department of Physics, Kennesaw State University, Marietta, GA, 30060, USA*

[2] *Materials Department and California Nano systems Institute, University of California Santa Barbara, Santa Barbara, California 93106, USA*

[3] *National High Magnetic Field Laboratory, Tallahassee, FL, 32310, USA*

[4] *Centro de Nanociencias y Nanotecnología, Universidad Nacional Autónoma de México, Ensenada, BC 22860, México*

[5] *Departamento de Física y Geociencias, Universidad del Norte, Barranquilla, Colombia*

**Abstract:**

The shape of the Fermi surface, and the cyclotron effective mass of the kagome magnet GdV$_6$Sn$_6$ charge carriers are investigated using de Haas van Alphen (dHvA) oscillations measurements and electronic band structure calculations. The temperature and angle-dependent torque magnetometry measurements revealed at least nine different frequencies ranging from ~10 T up to ~9000 T. These frequencies correspond to extremal areas of the Fermi surface ranging from ~0.2 % up to 50% of the first Brillouin zone, qualitatively consistent with the electronic band structure calculations. The angle dependent dHvA oscillation frequencies indicate that the smaller pockets of the Fermi surface have almost 3D character whereas the bigger pockets of the Fermi surface are mostly two-dimensional. We also find evidence of the presence of light (0.28(1) $m_0$) as well as heavy (2.37(18) $m_0$) charge carriers through the analysis of the temperature dependence of dominant frequencies. The comparison of the observed frequencies with the electronic band structure calculations indicates that the heavy masses correspond to saddle-point-like features of electronic band structure at the M point. The observation of the multiple low frequencies and the calculated contributions from various bands to such low frequencies prevent the estimation of topological nature of bands containing lighter fermions. In conclusion, our work reveals the features of a Fermi surface containing enhanced mass fermions originated from saddle points in the electronic band structure at the M point, which is inherent to kagome lattices.

**Introduction:**

The kagome lattice, a two-dimensional network of corner-sharing triangles of metal ions, is known to be a source of a variety of novel correlated electronic states [1–5]. The flat bands representing the correlated electronic states, the Dirac fermions featuring topological electronic states, and the saddle-point derived van Hove singularities causing novel electronic instabilities are typical features of kagome lattice materials [4]. Chiral charge density waves [6], Chern topological magnetism [7], and topological superconductivity [2,3,5] are some of the new electronic phases that have been observed in materials with kagome lattice structures.

The list of kagome metals includes chemically diverse compounds such as $Mn_3Sn$ [8], $Fe_3Sn_2$ [9,10], $Co_3Sn_2S_2$ [11–13], CoSn [14], FeSn [14], $AV_3Sb_5$ (A=Rb, Cs, K) [14], $RM_6X_6$ (R=Li/Mg/Yb/Sm/Gd/Ho/Tb/Y, M=Fe/Cr/Co/Ni/V, and X=Ge/Sn/Si) [7,15–22]. Such chemical diversity combined with layered crystal structures allows for fine tuning of intra- and inter-kagome layer interactions to realize novel electronic and magnetic phases. The family that draws particular attention is $RM_6X_6$. The $RM_6X_6$ structure contains two-dimensional parallel kagome layers of M ions coordinated by X ions and separated by the triangular planes of R ions. One advantage of such structure is that the inter kagome layer distances can be tuned by changing the size of R ions whereas the magnetic interactions can be varied by choosing the magnetic and non-magnetic R and M ions. Furthermore, the intrinsic physics associated with the kagome layer can be separated from the spacer layers by suitable choice of elements. The work presented in this manuscript is focused on the study of $GdV_6Sn_6$, in which the non-magnetic $V_3Sn_2$ kagome layers are separated by magnetic GdSn triangular planes and Sn atoms as shown in Fig. 1.

Previous studies on $GdV_6Sn_6$ indicate a non-collinear magnetic ground state ($T_N \sim 5$ K) arising from the *f*-orbitals of Gd ions along with a high mobility multiband electrical transport originating from the correlated electrons in the kagome layers [18–23]. The electronic band structure calculations as well as photoemission experiments indicate the presence of chemically-tunable Dirac surface states (DSS) [21,23], flat bands, and van Hove singularities featuring the intrinsic physics of kagome lattice [21–23]. Despite such studies, a detailed experimental investigation of the shape of the bulk Fermi surface, cyclotron effective mass of carriers, and observables featuring the topologically non-trivial bands and the saddle points causing van Hove singularity are still missing. Such experimental investigations can be carried out using de Haas van Alphen (dHvA) oscillations or Shubnikov- de Haas (SdH) oscillations. One previous study [19] of GdV$_6$Sn$_6$ uses SdH oscillations measurement of electrical resistivity. That previous study reports observations of two small frequencies 150 T and 200 T accounting for small Fermi pockets occupying about 2.5 % of the area of the first Brillouin zone. However, no other features revealing the relativistic nature of Dirac fermions and the saddle points in the electronic band structures were reported. In this work, we have used high field torque magnetometry measurements to study the dHvA oscillations. One advantage of the dHvA oscillations measurement is that the magnetization oscillation directly originates from the oscillations of the free electrons' energy and doesn't rely on scattering probabilities. By using a single crystalline sample of residual resistivity ratio (RRR ~ 12), we are able to observe dHvA oscillations on top of a magnetic background of ~ 7 $\mu_B$. We have extracted several oscillation frequencies ranging from 10 T- 9000 T indicating the presence of small and big pockets of the Fermi surfaces, consistent with the multiband nature of electrical transport and the calculated electronic band structure. The angular dependence of the oscillation frequencies indicates the presence of mostly 3D small pockets and quasi-2D type big pockets of the Fermi surfaces. The temperature dependence of the oscillation amplitudes indicates the presence of both light electrons (0.28 $m_0$) as well as heavier electrons (2.37 $m_0$). Some bands cross the Fermi level more than one time giving different effective masses. The observation of multiple low frequencies (<500 T), the calculated contributions from various bands, and crossing of the Fermi level by same band more than one time, prevent the estimation of the Berry curvature associated to topologically non-trivial bands. However, we are able to clearly observe and resolve other features associated with enhanced mass fermions that characterize kagome materials such as the saddle points in the proximity of the M-point in the Brillouin zone.

**Experimental Details:**

Single crystals of GdV$_6$Sn$_6$ were synthesized via a flux-based technique. Gd (pieces, 99.9%), V (pieces, 99.7%), Sn (shot, 99.99%) were loaded inside an alumina crucible with the molar ratio of 1:6:20 and then heated at 1125 °C for 12 hours. Then, the mixture was cooled at a rate of 2 °C/h. The single crystals were separated from the flux via centrifuging at 780 °C. Crystals grown via this method were generally a few millimeters long and <1 mm in thickness. The separated single crystals were subsequently cleaned with dilute HCl to remove any flux contamination. Crystals were then transferred into a small jar of mercury to further remove additional tin contamination to the crystals. Single-crystal x-ray diffraction measurements were carried out on a Kappa-Apex II single crystal diffractometer with a charge coupled device (CCD) detector and a Mo source. The low field magnetization measurements were carried out using a Quantum Design Magnetic Properties Measurement Systems (MPMS-3). The resistivity was measured using four probe methods employing the electrical transport option (ETO) of the Quantum Design Dynacool Physical Properties Measurement System.

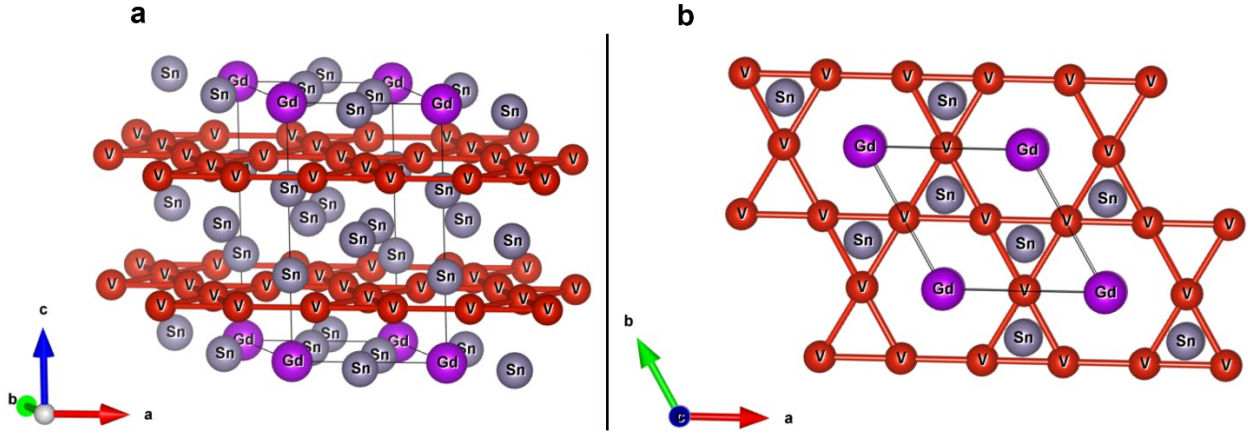

**Figure 1**: Crystal structure of $GdV_6Sn_6$. (a) Crystal structure showing different layers of Gd, Sn and V atoms. (b) Crystal structure viewed along the *c*-axis showing the kagome network of V atoms. The colored spheres indicate different atoms.

High-field measurements were carried out at the National High Magnetic Field Laboratory (NHMFL), Tallahassee, Florida, with the maximum applied fields of 18 T (Superconducting magnet), and 35 T (dc resistive water-cooled magnet). In both experiments the lowest temperature of 0.35 K was achieved using a top-loaded $^3$He insert. The magnetic torque was measured using a miniature piezoresistive cantilever. A tiny $GdV_6Sn_6$ crystal was selected and then fixed to the cantilever arm with vacuum grease. The cantilever was subsequently mounted on the rotating platform of a special probe designed at NHMFL. The probe was then slowly cooled down to the base temperature of 0.35 K. Two resistive elements on the cantilever were incorporated with two other room-temperature resistors to form a Wheatstone bridge, which was balanced at base temperature before taking field dependent data. The angle-dependent torque data were obtained by rotating the sample in situ with the applied field. Magnetic fields were swept at each fixed temperature at a rate of 2.7 T/min (up) and 4.2 T/min (down).

**Computational methods:**

The electronic band structure calculations were done using the Vienna Ab Initio Simulation Package (VASP) [24–26]. The electron-electron non-classical exchange-correlation interactions were modeled using the generalized gradient approximation under the PBE parametrization [27]. Projected augmented wave potentials [25,28] were used with an optimized cutoff energy of 520 eV. An energy and force criteria were applied to reach structural relaxation, where the energy and norms of all forces must be less than $10^{-6}$ eV and 0.01 eV/Å, respectively. A discrete equally spaced mesh [29] of 10×10×6 k-points was used to evaluate the electronic states during the ionic relaxation process. For the electronic band structure analysis, the convergence criterion was taken as the condition where the total energy change between two successive iterations in the self-consistent loop became smaller than $10^{-8}$ eV. For the converged calculations, a k-point mesh of 15×15×12 was used. Since we are treating atoms with large atomic mass -where relativistic effects take importance- spin-orbit coupling interactions are mandatory, we also included it in the relaxation and electronic properties of the GdV$_6$Sn$_6$ structure. Also, *f*-orbitals of the Gd atom with highly localized electrons must be accounted for. To do so, we included the Hubbard interaction in the simplified approach proposed by Dudarev et al. [30], with an on-site Coulomb parameter U= 6 eV for the Gd atom. To calculate the Fermi energy in a dense k-mesh (121×121×123), we have used the Hamiltonian based on the Wannier functions obtained using the Wannier90 code [31]. We modeled the GdV$_6$Sn$_6$ material considering the hexagonal P6/mmm space group in its ferromagnetic structure. The dHvA frequencies and their angular dependencies were calculated via the skeaf code [32]

**Results:**

The room temperature X-ray diffraction pattern from the flat surface of a single crystal of GdV$_6$Sn$_6$ is presented in Fig. 2a. The peaks can be indexed with hexagonal structure with space group P6/mmm. The diffraction pattern contains only peaks corresponding to Miller indices (00*l*, l=1,2, 3..) indicating that the flat surface is perpendicular to crystalline *c*-axis. The magnetic susceptibility of a single crystal under a field of 1 *kOe* applied along the *c*-axis is presented in Fig. 2b. The susceptibility follows a typical Curie-Weiss behavior at higher temperatures; however, it enters a long-range magnetic phase around $T_N \approx 5$ K. The inset in Fig. 2b displays the fitting to Curie-Weiss behavior $\frac{1}{\chi} = \chi_0 + \frac{C}{T-\theta_C}$ (T >20 K) with the Curie-constant (*C*), a Curie-Weiss temperature ($\theta_C$), and a small background term ($\chi_0$). The fitting yields $C$=7.307(2) emu mol$^{-1}$ Oe$^{-1}$ K, $\theta_C$ = 7.96 (2) K and $\chi_0$ = 0.002(1) emu mol$^{-1}$ Oe$^{-1}$ K. This gives effective moment $\mu_{eff}$ = 7.7(2) $\mu_B$/f.u consistent with the expected full moment from Gd$^{3+}$ ion (7.94 $\mu_B$). The positive Curie-Weiss temperature indicates the dominant ferromagnetic interactions along the $c$-axis. Figure 2c presents the zero-field electrical resistivity as a function of temperature for a GdV$_6$Sn$_6$ single crystal with current within the $ab$-plane. The resistivity exhibits typical metallic behavior with a small downturn at the magnetic transition (5 K), indicating the interaction between 3$d$-itinerant electrons in the kagome layer and the Gd spins in the spacer layer. The residual resistivity ratio (RRR ≈ 12) allows the measurement of the dHvA oscillations in the presence of uniform magnetic background. A more detailed investigation of the structural, electrical, and magnetic properties of GdV$_6$Sn$_6$ indicating the onset of long-range magnetic order with a large, saturated moment (~7 $\mu_B$), consistent with both our computational findings (7.07 $\mu_B$ per Gd ion) and the multiband behavior observed in previous electrical transport studies [22].

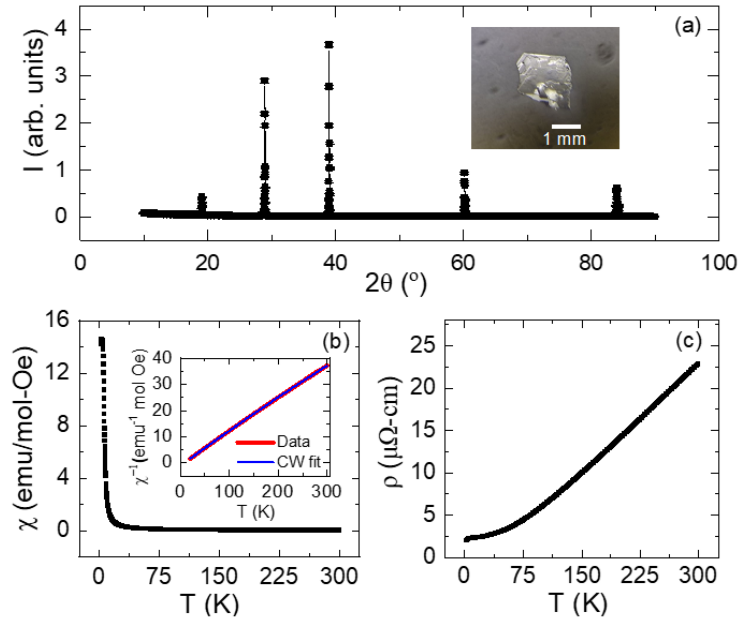

**Figure 2**: Single crystal characterization of GdV$_6$Sn$_6$. (a) X- ray diffraction pattern observed from the flat surface of a single crystal (shown in inset) of GdV$_6$Sn$_6$. The presence of only sharp (*00l*) type reflections indicate high quality single domain crystal with the $c$-axis perpendicular to the flat surface. (b) Magnetic susceptibility, $\chi$, as a function of temperature measured at 1 *kOe* field applied parallel to the $c$-axis. The inset shows Curie Weiss law fitting for $T$ > 20 K. (c) Zero field electrical resistivity, $\rho$, with current within the $ab$ plane.

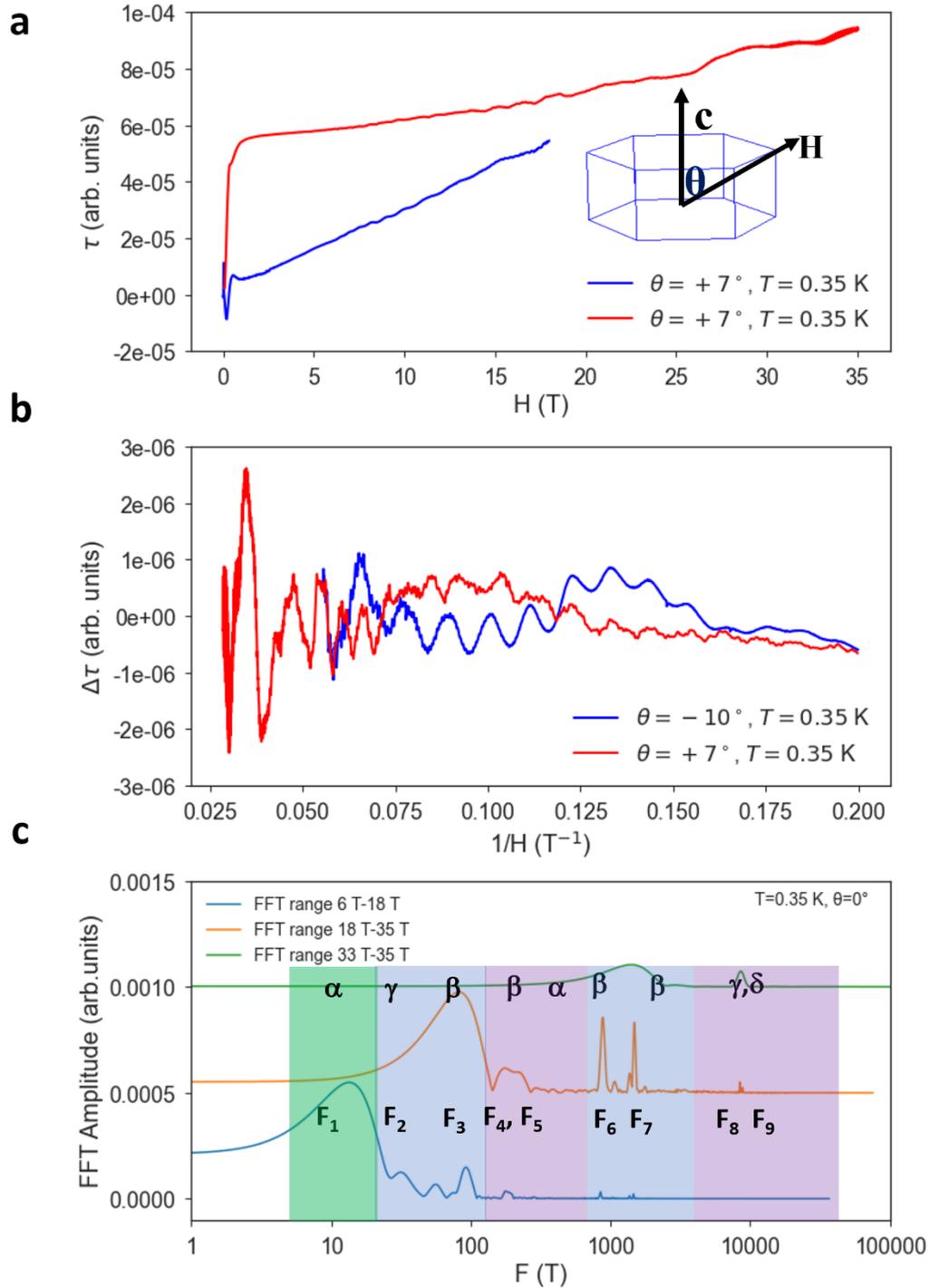

**Figure 3:** Magnetic torque measurements (a) The magnetic torque ($\tau$) as a function of magnetic field (H) taken using the 18 T (blue) and the 35 T (red) magnet at given angles. The inset figure shows the schematic of the magnetic field direction relative to crystal axes. (b) The polynomial background subtracted signal ($\Delta\tau$) as a function of the inverse magnetic field (1/H) (c) The fast Fourier transform (FFT) spectra (amplitude vs frequency) at $\theta = 0°$, T= 0.35 K, for FFT range of 6 T-18 T, 18 T- 35 T, 33 T- 35 T. The x-axis is plotted in logarithmic scale to highlight features at low frequencies. The frequencies $F_1, F_2, F_3, F_4, F_5, F_6, F_7, F_8, F_9$ in Fig. (c) represent dominant frequencies. The symbols ($\alpha, \beta, \gamma, \delta$) represent possible bands contributing to those frequencies as indicated by shaded regions.

The results of typical magnetic torque measurements are presented in Fig. 3. Fig. 3a presents the magnetic torque ($\tau$) as a function of magnetic field ($H$) at $\theta$ = -10° and + 7° measured using the 18 T superconducting magnet and the 35 T resistive magnets respectively. Here, $\theta$ is the angle between the magnetic field and the *c*-axis of the crystal as shown in inset of Fig. 3a and the sign of $\theta$ represents the sense of rotation of the magnetic field with respect to the *c* axis of the crystal. Fig. 3b displays the third order polynomial background subtracted signal ($\Delta\tau$) as function of inverse magnetic field ($1/H$) at given angles indicating quantum oscillations that are periodic in $1/H$. Fig. 3c displays the Fourier transformation of background subtracted signals at $\theta$ = 0°, $T$ = 0.35 K for different Fast Fourier Transformation (FFT) ranges.

The *x*-axis of Fig. 3c is displayed in a logarithmic scale to highlight the low frequencies. At $\theta$ = 0°, we observed 9 dominant frequencies, $F_1$ = (13 ± 7) T, $F_2$ = (32 ± 10) T, $F_3$ = (95 ± 10) T, $F_4$ = (176 ± 6) T, $F_5$ = (200 ± 10) T, $F_6$ = (850 ± 20)T, $F_7$ = (1470 ± 30 )T, $F_8$ = (8440 ± 50) T, $F_9$ = (8850 ± 60) T, that are consistent with the electronic band structure calculations. These frequencies correspond to four different bands ($\alpha$, $\beta$, $\gamma$, and $\delta$) present at the Fermi level. We also observed frequencies having weak amplitudes around 55 T, 290 T, 390 T, 1040 T, and 1370 T. However, due to the presence of nearby frequencies with strong amplitudes, we are unable to track down the temperature and angle dependence of all frequencies. We therefore focus on the analysis of 6 frequencies ($F_1$, $F_3$, $F_6$, $F_7$, $F_8$, $F_9$) mentioned above and displayed in Fig. 3c. After identifying the frequencies at $\theta$ = 0°, the angle-dependent torque measurements are performed to understand the shape of the Fermi surface. Figure 4 displays the results of angle-dependent measurements at $T$ = 0.35 K. For the analysis of low frequencies (F < 500 T), FFT range of 6 T- 18 T is used whereas for frequencies F > 500 T, the FFT range of 18 T- 35 T is used. As is clear from Fig. 3c and Fig. 4, we observed many frequencies below 500 T making it difficult to follow the angle dependence of every frequency. The low frequencies $F_1$, $F_2$, $F_3$, $F_4$, $F_5$ survive at all angles between (0 to -90) with some angle dependence. The frequencies $F_6$ and $F_7$ appear only between $\theta$ = 0° to $\theta$ = -10° and disappear quickly. The high frequencies $F_8$, $F_9$ disappear above $\theta$ = -50°. This indicates that the Fermi surface contains small pockets of mostly 3D shape and bigger pockets of 2D shapes. It is important to note that out of these frequencies observed in this work, only two nearby frequencies of 150 T and 200 T were reported in the previous work [19] using SdH oscillations.

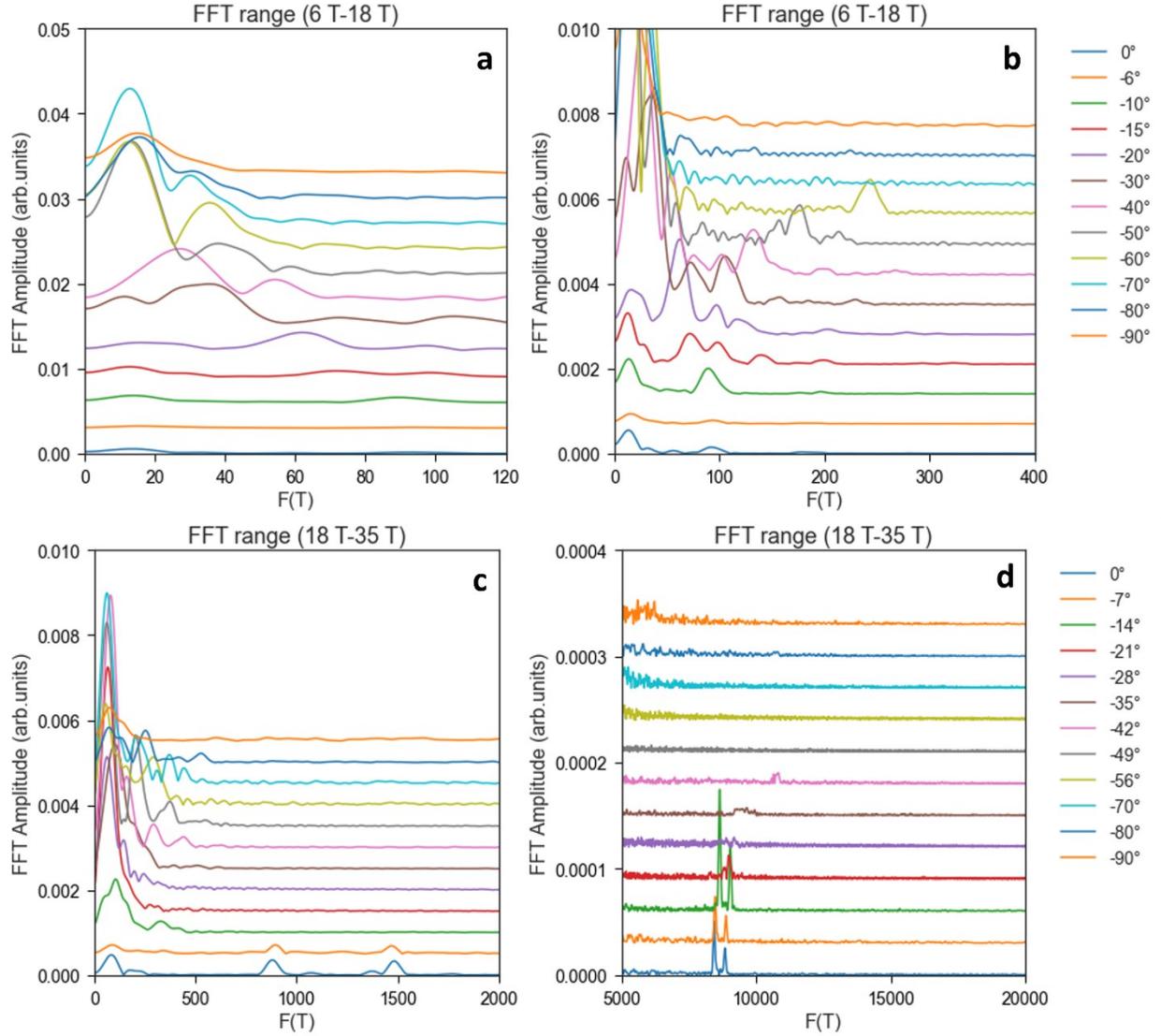

**Figure 4:** Angle dependence of dHvA oscillation frequencies at T = 0.35 K. (a) & (b) Frequencies obtained with FFT range of 6 T- 18 T. (c) & (d) Frequencies obtained with FFT range of 18 T -35 T.

The observed frequencies are compared to those expected from electronic band structure calculations. The results are summarized in Fig. 5. The electronic-structure-calculated frequencies from the different bands are plotted as colored symbols (▲,▼,♦,◀)whereas the observed frequencies are plotted as (●). The calculated frequencies for the four bands ($\alpha$, $\beta$, $\gamma$, and $\delta$) closely match those observed experimentally. The next step is to study the temperature dependence of these frequencies. By suitably choosing the FFT range and the angle, the temperature dependence of $F_1$, $F_3$, $F_6$, $F_7$, $F_8$ and $F_9$ are studied in this work. The temperature dependence of $F_1$, $F_3$, $F_8$ and $F_9$ are studied using data taken at $\theta$ = -10° whereas the temperature dependence of $F_6$ and $F_7$ are studied using data taken at $\theta = 0°$. We used an FFT range of 6 T -18 T to study the temperature variations of $F_1$, $F_3$ and $F_5$ whereas we used FFT range of 14 T - 18 T to study the temperature variation of $F_6$ and $F_7$. The variation of normalized FFT amplitudes with temperature for $F_1$, $F_3$, $F_6$, $F_7$, $F_8$, and $F_9$ are presented in Figure 6. These variations can be described by the damping part of the Lifshitz-Kosevich (LK) formula [33,34] (solid lines in Figure 6). Fitting with the LK formula gives six different effective masses $m_1^* = 0.28(1)\ m_0$, $m_3^* = 0.36(1)\ m_0$, $m_6^* = 1.20(6)\ m_0$, and $m_7^* = 0.79(5)\ m_0$, $m_8^* = 2.25(15)\ m_0$, and $m_9^* = 2.37(18)\ m_0$. The details of the LK fitting are presented in Appendix B.

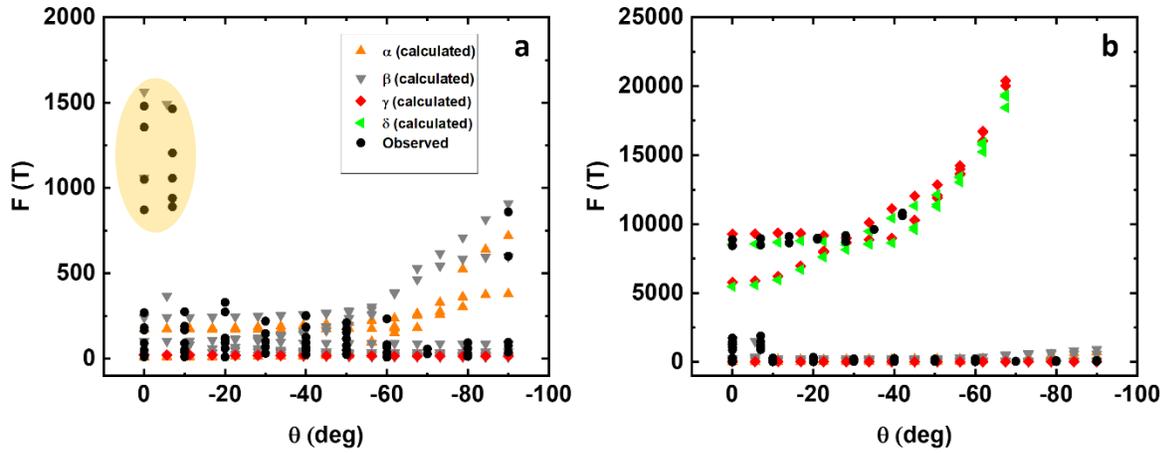

**Figure 5:** (a) Angle dependence of electronic band structure calculated frequencies (α, β, γ, δ) and observed frequencies (●). The observed frequencies are taken at $T=0.35$ K. Different colors of calculated frequencies represent different bands. (a) Observed and calculated frequencies below 2000 T. The shaded region in (a) contains at least two frequencies ($F_6$ and $F_7$) related to two orbits containing β-band. The other frequencies in (a) contain contributions from at least 3 different orbits associated to α, β, and γ bands. (b) All observed and calculated frequencies. The frequencies above 5000 T in (b) correspond to γ and δ bands ($F_8$, $F_9$) respectively.

Along with the determination of the effective mass ($m^*$) using the LK formula, the different areas associated with different sections of the Fermi surface are estimated using the Onsager relation [34]. We then calculated the Dingle temperature $T_D$ (an additional temperature factor that accounts for the damping of oscillations amplitude with inverse field). The estimation of Dingle temperature is presented in Appendix B. After calculating the extremal area ($S_f$), Fermi wave vector ($k_f$), effective mass ($m^*$), and Dingle temperature ($T_D$), we have estimated the Fermi velocity ($v_f$), quantum scattering time ($\tau_s$), mean free path ($l_D$), and the quantum mobility ($\mu$). We define $\mu$ as quantum mobility to distinguish from the classical mobility arising from the Drude model that can be present without magnetic field. In this case, mobility is related to the cyclotron motion of carriers. These quantities are presented in Table I.

**Table I:** Results of dHvA oscillations showing observed frequencies (F), extremal orbit area ($S_f$), Fermi wave vector ($k_f$), effective mass ($m^*$), Fermi velocity ($v_f$), Dingle temperature ($T_D$), quantum scattering time ($\tau_s$), mean free path ($l_D$), and quantum mobility ($\mu$). The numbers on bracket indicate the errors.

| F (T) | $S_f$ (Å$^{-2}$) | $k_f$ (Å$^{-1}$) | $m^*/m_0$ | $v_f$ (10$^4$ ms$^{-1}$) | $T_D$ (K) | $\tau_s$ (10$^{-13}$ s) | $l_D$ (nm) | $\mu$ (cm$^2$ V$^{-1}$ s$^{-1}$) |
|---|---|---|---|---|---|---|---|---|
| 95 (10) | 0.009(1) | 0.053(2) | 0.36(1) | 17(1) | 13.1(2) | 0.92(1) | 16(1) | 453(14) |
| 850 (20) | 0.081(1) | 0.161(1) | 1.21(1) | 15.3(2) | 10.2(3) | 1.19(3) | 18(1) | 173(5) |
| 1470 (30) | 0.141(2) | 0.211(2) | 0.79(5) | 31(2) | 22(1) | 0.55(2) | 17(1) | 123(9) |
| 8440 (50) | 0.802(4) | 0.506(1) | 2.25(15) | 26(2) | 4.1(2) | 2.9(1) | 77(4) | 231(19) |

After establishing the presence of multiple pockets of the Fermi surface, the next conventional step in such analysis would be the estimation of Berry phase ($\phi_B$) by analyzing the oscillatory part of the LK formula [34,35]: $\sin\left[2\pi\left(\frac{F}{<H>} + \frac{\varphi_B}{2\pi} - \frac{1}{2} + \delta_p\right)\right]$ where, the factor $\delta_p$ depends upon the dimensionality of the Fermi pocket and takes value (+1/8) for the minimal and (-1/8) for maximal cross sections of 3 dimensional Fermi surface. The factor of ½ comes from Maslov correction and applies to the case where the orbits are compressible to circles [36].

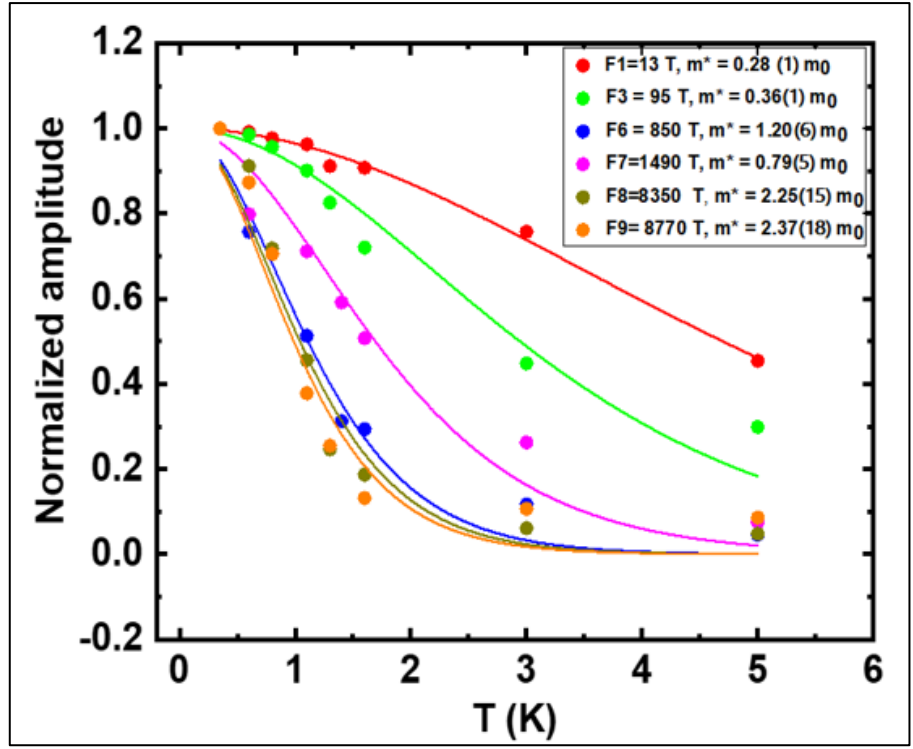

**Figure 6:** Temperature dependence of amplitudes of dominant frequencies. Data for $F_6$ and $F_7$ were taken from $\theta = 0$ measurements while the rest data were taken from $\theta = -10^0$. The solid lines fit to the damping part of LK formula described in Appendix B.

It has been suggested in recent works [35,37,38] that the precise estimation of Berry phase requires understanding the details of crystalline and magnetic symmetries of materials. Therefore, a phase of π should not be taken as smoking gun proof of non-trivial topology. In our experimental study, we encountered a notable complication stemming from the presence of numerous closely spaced frequencies, all falling below the 500 T threshold. This intricate frequency landscape presents a significant challenge, rendering the precise estimation of the Berry phase inconclusive. However, unlike the Berry phase that could be blurred by multiple close frequencies, the mass enhancement associated to the saddle points in electronic band structure are clearly observed and reproduced by electronic band structure calculations.

**Discussion and conclusions**: Our dHvA-based analysis of $GdV_6Sn_6$ revealed several important features revealing this kagome material's Fermi surface properties. We observed 9 dominant frequencies [$F_1$= (13 ± 7) T, $F_2$ = (32 ± 10) T, $F_3$ = (95 ± 10) T, $F_4$ = (176 ± 6) T, $F_5$= (200 ± 10) T, $F_6$ = (850 ± 20)T, $F_7$= (1470 ± 30 )T, $F_8$ = (8440 ± 50) T, $F_9$= (8850 ± 60) T ], when the magnetic field is applied perpendicular to the kagome plane (θ = 0°). These frequencies are consistent with electronic band structure calculations. In addition, we observed frequencies with weak amplitudes around 55 T, 290 T, 390 T, 1040 T, and 1370 T. However, due to the presence of nearby strong amplitude frequencies, we are unable to follow the temperature and angle dependence of all frequencies. The lowest frequency $F_1$ correspond to less than 2 periods of oscillation in the FFT range of 6 T-18 T, but it shows a discernible temperature dependence close to the calculated frequencies from α, and γ bands. The comparison of the observed frequencies with the calculated frequencies indicates that the frequencies $F_1$ and $F_4$ are most likely related to orbits orbit originating from the α band, $F_2$ and $F_8$ related to orbits originating from the γ band, $F_3$, $F_5$, $F_6$, and $F_7$ from the β band and $F_9$ from the δ band. Furthermore, the angle dependence of dHvA frequencies indicates that the low (F < 500 T) frequencies survive when the magnetic field is rotated θ = 0° to θ = 90°, indicating small Fermi pockets of mostly 3D shape. We also observed large pockets of the Fermi surface. The two frequencies related to such large Fermi pockets disappear at high angles indicating their quasi-2D nature. These facts are also supported by the calculated frequencies from electronic band structure calculation (Fig.5 and Fig.7) and the projected Fermi surface presented in Fig. 7b. Notably, α and β bands exhibit hole-like characteristics, while the γ and δ bands predominantly demonstrate electron-like behavior. Regarding the Fermi surface shapes, the α and β bands exhibit irregular lobes or pockets along the M-L high symmetry path whereas, the bands γ and δ display barrel-like shapes with prominent and open regions along the same M-L high symmetry path, as illustrated in Figure 7a and 7b. It is to be noted that the FFT range for angle dependence presented here is (6 T-18 T) for low frequencies (F<500 T) and (18 T-35 T) for high frequencies (F > 500 T). While changing the FFT window, we are also able to observe (not presented here) the second harmonics at some angles especially for high frequencies (F>500 T). Furthermore, from the temperature dependence of amplitudes of dominant frequencies, we have estimated the cyclotron effective masses [$m_1^* = 0.28(1)\, m_0$, $m_3^* = 0.36(1)\, m_0$, $m_6^* = 1.20(6)\, m_0$, and $m_7^* = 0.79(5)\, m_0$, $m_8^* = 2.25(15)\, m_0$, and $m_9^* = 2.37(18)\, m_0$]. The light masses correspond to α and β bands whereas the heavy masses ($m_8$, and $m_9$) correspond to relatively flat bands (γ and δ) near saddle point around the M point. Notably, these enhanced mass fermions correspond to vanadium d-orbitals near van Hove singularity ($VHS_2$) that is clear from the density of states plot presented in Figure 8 b. It is also observed that the calculated quantum mobilities are relatively small compared to other non-magnetic kagome materials such as $CsV_3Sb_5$ [39,40]. This is most likely caused by the scattering of electrons from the magnetic background.

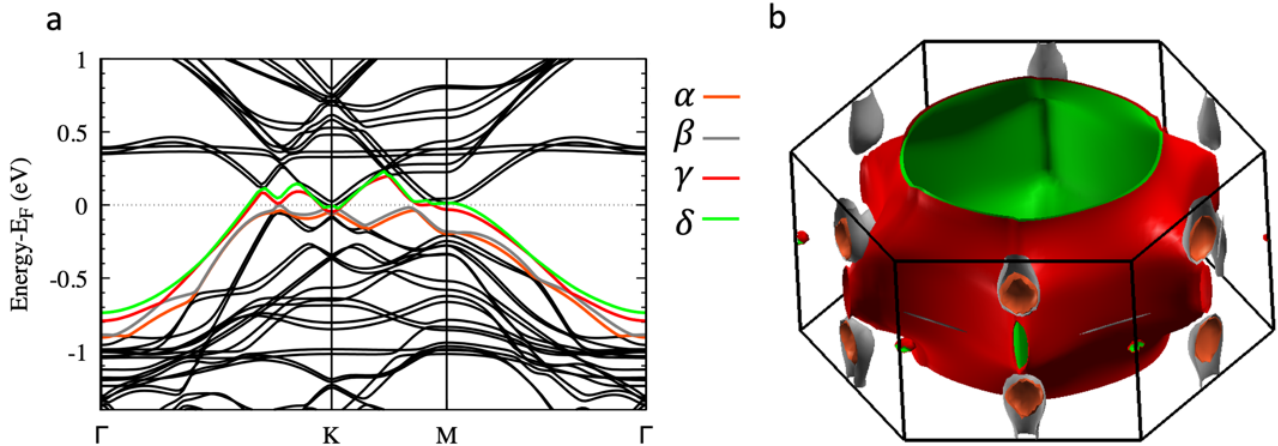

**Figure 7:** (a) The electronic band structure of $GdV_6Sn_6$ in the ferromagnetic state; (b) Fermi surface representation illustrating distinct pockets. The four electronic bands (α, β, γ, and δ) crossing the Fermi level are depicted in distinct colors.

Finally, the electronic band structure presented in the current work (Fig. 7 and Fig. 8) is in the ferromagnetic state, but it remains largely unchanged in the vicinity of Fermi level when compared to electronic band structure calculated in the paramagnetic state of TbV$_6$Sn$_6$ [41] and GdV$_6$Sn$_6$ [22].

This stability is attributed to the f-orbitals, crucial for the material's magnetism, being located away from the Fermi level. Moreover, the topological features and saddle points around the Fermi level are dominated by the kagome layer of V atoms.

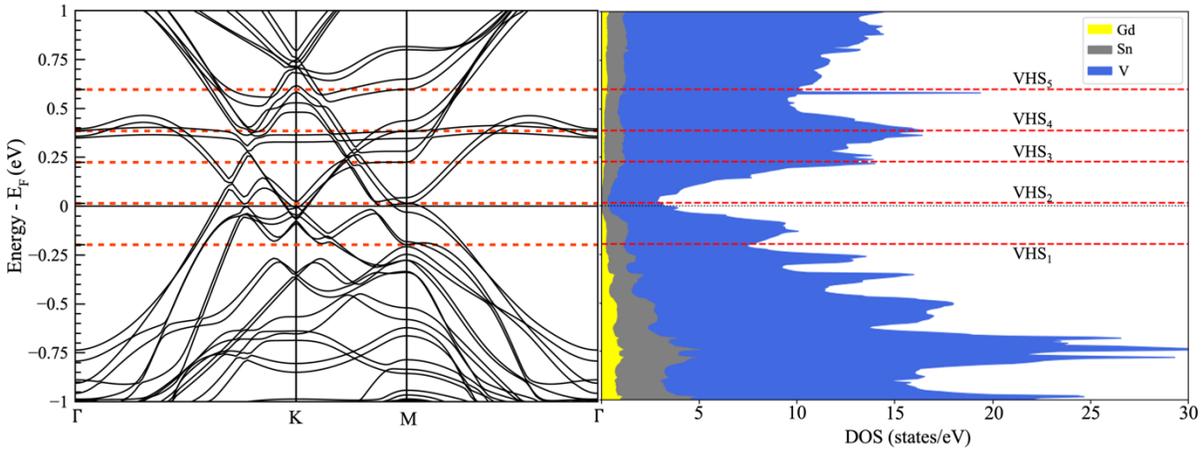

**Figure 8.** Band structure of ferromagnetic GdV$_6$Sn$_6$ including SOC and projected density of states for the Gd-f orbitals (yellow), Sn-p orbitals (gray) and V-d orbitals (blue). Fermi level is set to zero energy. Dashed lines indicate the energy appearance of the Van Hove Singularities (VHS).

In summary, our work reveals the Fermi surface of kagome magnet GdV$_6$Sn$_6$ with small and big pockets of the Fermi surface consistent with the multiband nature of electrical transport measurements [22]. We found the existence of both lighter (0.28 m$_o$) and heavier (2.37 m$_o$) fermions. While the topological nature of bands contributing to light mass is uncertain, we found clear evidence of enhanced mass fermions originating from saddle point like feature of electronic band structure at the proximity of the M point that led to VHS in the density of states at the Fermi level. The appearance of such saddle point and corresponding VHS (VHS$_2$) in the proximity of Fermi level implies that this system is susceptible to various electronic instabilities.

Note: After this manuscript was submitted for review, we became aware of the similar Fermi surfaces with light and heavy fermions in sister material YV$_6$Sn$_6$ [42] reflecting the features of topological non-trivial bands and saddle points like features of electronic band structure of kagome materials.

**Acknowledgments**: This work is based upon the work supported by the National Science Foundation under grant number DMR- 2213443. We thank J.F DiTusa for the illuminating discussions. GP and SDW acknowledge the support via the UC Santa Barbara NSF Quantum Foundry funded via the Q-AMASE-i program under award DMR-1906325. A portion of this work was performed at the National High Magnetic Field Laboratory, which is supported by the National Science Foundation Cooperative Agreement No. DMR-1644779 and the State of Florida. J.G.S acknowledges DGAPA-UNAM project IG101124 for partial financial support. Calculations were performed in the DGCTIC-UNAM Supercomputing Center project LANCADUNAM-DGTIC-368, LNS-BUAP project 202201042N, and THUBAT KAAL IPICYT supercomputing center project TKII-JGSA001. J.G.S. also acknowledges A. Rodriguez-Guerrero and E. Murillo for useful discussions and technical assistance. M. M. A. is grateful for the support of the National Science Foundation through grant number DMR-2213429.

**Appendix A: Angle dependence around θ = 0°**

The angle dependencies of FFT frequencies around θ = 0° are presented in Fig. 9a and Fig.9b. The data was collected using 18 T superconducting magnets at intervals of θ = 3° to 6°.

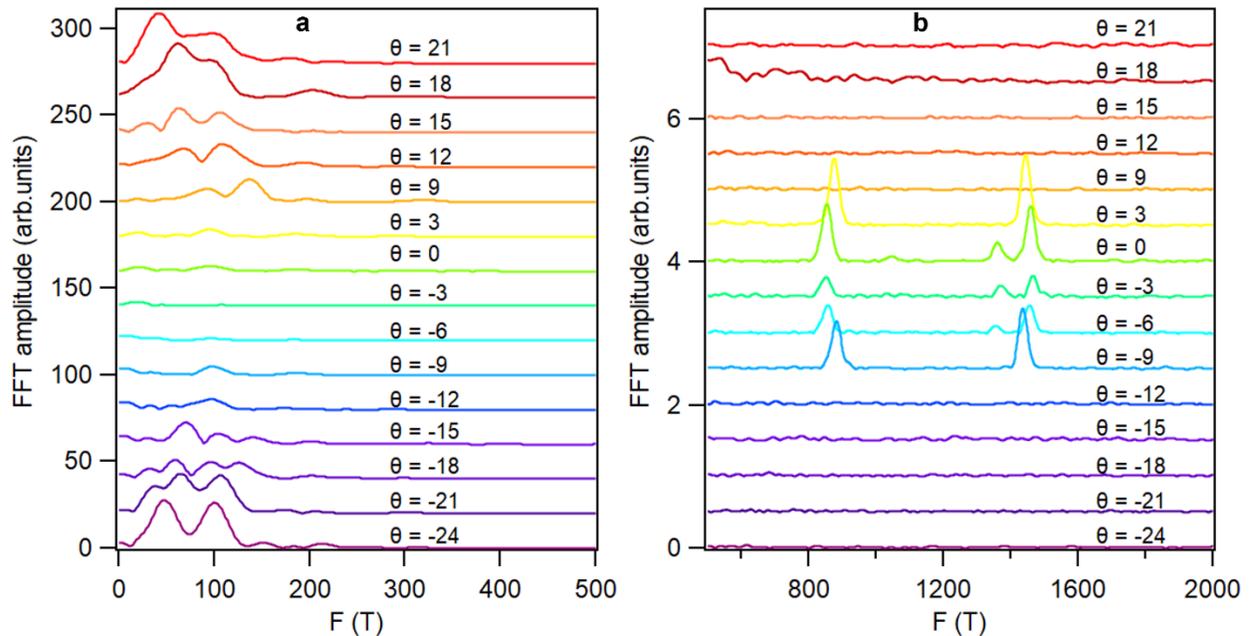

**Figure 9:** Small steps angle dependence around θ=0° at T= 0.35 K (a) Angle dependence of FFT amplitudes for frequency range 0-500 T (b) angle dependence of FFT amplitudes for frequency range 500-2000 T. The angle (θ) is expressed in degrees (°).

**Appendix B: Lifshitz- Kosevich (LK) formula, Dingle temperature ($T_D$), Onsager relation, Fermi velocity ($v_f$), scattering rate ($\tau_s$), mean free path ($l_D$), and quantum mobility ($\mu$)**

The oscillatory part of torque is given by [34,43,44]

$$\Delta\tau \propto H^\lambda \frac{A\left(\frac{m^*}{m_0}\right)\frac{T}{<H>}}{\sinh\left(A\left(\frac{m^*}{m_0}\right)\frac{T}{<H>}\right)} \exp\left\{-A\left(\frac{m^*}{m_0}\right)\frac{T_D}{<H>}\right\} \cos\left(\pi g \frac{m^*}{2m_0}\right) \sin[2\pi(\frac{F}{<H>} + \psi)],$$

where $\frac{A\left(\frac{m^*}{m_0}\right)\frac{T}{<H>}}{\sinh\left(A\left(\frac{m^*}{m_0}\right)\frac{T}{<H>}\right)}$ is thermal damping factor, $\exp\left\{-A\left(\frac{m^*}{m_0}\right)\frac{T_D}{<H>}\right\}$ is the Dingle damping factor, $T_D$ is the Dingle temperature, $\cos\left(\pi g \frac{m^*}{2m_0}\right)$ is spin reduction factor, g is Lande "g" factor, $m^*$ is effective mass of electrons (holes), $m_0$ is mass of free electron, and the exponent $\lambda \sim 0$ for 2D Fermi surface, and $\lambda \sim \frac{1}{2}$ for 3D Fermi surface [34]. The constant A is given by: $A = \frac{2\pi^2 k_B m_0}{e\hbar} = 14.69$ T/K. Here $<H>$ is the harmonic mean of minimum and maximum field used in FFT. $[\frac{1}{<H>} = \frac{\left(\frac{1}{H_{min}} + \frac{1}{H_{max}}\right)}{2}]$. The phase factor $\psi$ is given by $\psi = \left(\varphi - \frac{1}{2}\right)$. The factor $\varphi$ is given by $\varphi = \frac{\emptyset_B}{2\pi} + \delta_p$. Here $\emptyset_B$ is the Berry phase and $\delta_p$ is 0 for 2D and $\pm 1/8$ for 3D Fermi surfaces with minimal and maximal cross sections respectively [44]. The effective mass ($m^*$) is calculated by fitting the normalized amplitude of oscillations to thermal damping factor term $\frac{A\left(\frac{m^*}{m_0}\right)\frac{T}{<H>}}{\sinh\left(A\left(\frac{m^*}{m_0}\right)\frac{T}{<H>}\right)}$. The Dingle temperature $T_D$ is obtained by fitting the Dingle damping factor term $[\exp\left\{-A\left(\frac{m^*}{m_0}\right)\frac{T_D}{<H>}\right\}]$. In practice this is done by finding the slope of $\ln\left[\Delta\tau H^{0.5} \sinh\left(A\left(\frac{m^*}{m_0}\right)\frac{T}{<H>}\right)\right]$ vs $1/H$ plot and dividing the slope by $A\left(\frac{m^*}{m_0}\right)$ factor (Figure 10). For the calculation of extremal area of Fermi surface, we have used the Onsager relation: $F = \left(\frac{\emptyset_0}{2\pi^2}\right) S_f$, here $\emptyset_0 = 2\pi\hbar/e$, is the flux quantum, and $S_f = \pi k_f^2$ is the extremal area of Fermi surface normal to the magnetic field. Here $k_f$ is Fermi wavevector. The Fermi velocity is calculated using $v_f = \frac{\hbar k_f}{m^*}$. The scattering rate ($\tau_s$) is calculated from Dingle temperature using the relation: $\tau_S = \frac{\hbar}{2\pi k_B T_D}$, the mean free path is given by $l_D = v_f \tau_S$, and the mobility is calculated using relation $\mu = \frac{e\tau}{m^*}$.

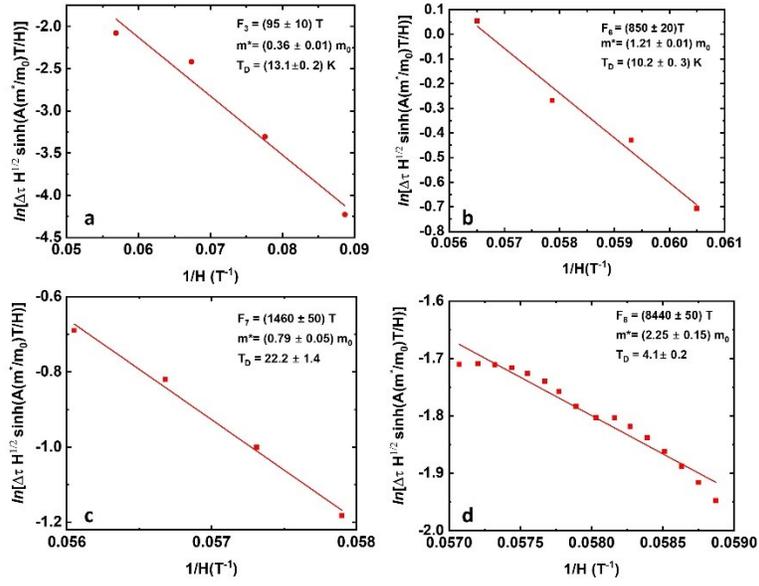

**Figure 10**: Dingle temperature calculation for (a) Frequency $F_3$ (b) $F_6$ (c) $F_7$, and (d) $F_8$.

## Appendix C: Electronic band structure calculation

We modeled the GdV$_6$Sn$_6$ material considering the hexagonal P6/mmm space group in its ferromagnetic structure. The optimized lattice parameters a = 5.518 Å and c = 9.265 Å that are close to the experimental values a = 5.5348(7), c = 9.1797(11) Å [22]. The obtained magnetic moments in the V-d and Gd-f orbitals are -0.147 μ$_B$ and -6.928 μ$_B$, respectively. This indicates that the Gd-f states dominate the ferromagnetic state in this material.

After optimization of the structure, we calculate the electronic band structure through band structure and projected density of states. Figure 11a depicts the band structure at the high symmetry points in the irreducible Brillouin zone. The band structure depicts the well-known flat bands induced by the kagome structure (~0.35 eV), mainly due to the Vanadium-d orbitals [22]. Near the Fermi level, Dirac-like dispersion relations are observed at the K symmetry point due to the hexagonal symmetry of the Kagome lattice. We also note that bands have some linear dispersion at the M point and in the Γ-K path. The band structure evidences an apparent metallic anisotropic behavior with a large band gap energy at the Γ-A path but conduction states along the L-M (Figure 11b) and H-K paths.

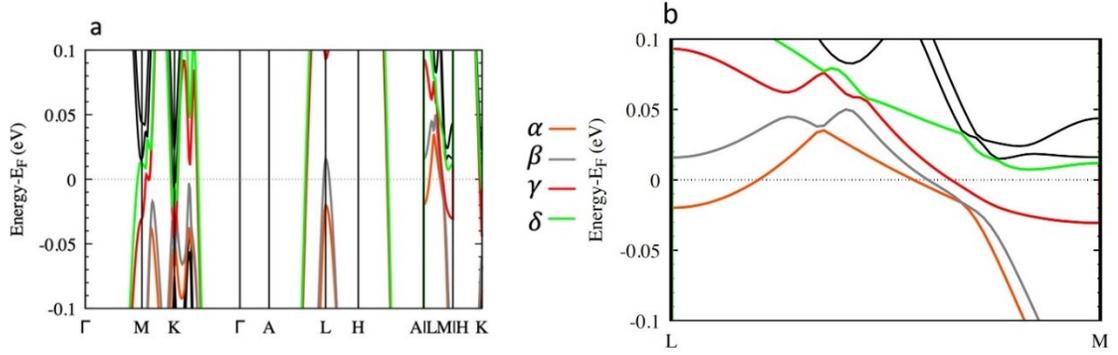

**Figure 11.** (a) Band structure of ferromagnetic GdV$_6$Sn$_6$ and (b) Band structure along L-M line of the Brillouin zone, zoomed in from +100meV to -100meV. The four electronic bands (α, β, γ, and δ) crossing the Fermi level are depicted in distinct colors.

In Figure 8, we plot the band structure and the projected density of states to evidence the appearance of multiple Van Hove Singularities (VHS) near the Fermi level. In the density of states, VHSs appear as sharp changes like peaks, valleys, or cusp-like structures, while in the band structure, they appear as local extrema such as saddle points, points with large curvature, or band edges. Both characteristics coincide in energy, as shown in Figure 8. Previous DFT calculations without including spin-orbit calculations evidenced the appearance of four VHSs which were labeled as VHS$_1$, VHS$_2$, VHS$_3$, and VHS$_4$ [21], in agreement with our findings, in which we have included the SOC effect and considered the Ferromagnetic phase. As expected, the SOC effect generates band splitting and a potential change in the VHS form, energy of appearance, and quantity. For example, we also observed one more VHS$_5$ and potentially another at -0.5 eV. All these points appear at the M high symmetry point. Our SOC calculations evidence that the VHSs are mainly formed by the d-orbitals of the vanadium kagome lattice.

## Appendix D: Fermi surface

We have obtained the Fermi Surface through first-principles calculation, employing the optimized Ferromagnetic structure. In this analysis, we have discerned four distinct electronic bands: α, β, γ, and δ. The Fermi surfaces have been projected at the Fermi energy, indicating that all observed bands intersect with this critical energy level. Notably, Band-α and Band-β exhibit hole-like characteristics, while Band-γ and Band-δ predominantly demonstrate electron-like behavior. Regarding the Fermi surface shapes, Bands α and β exhibit irregular lobes or pockets along the M-L high symmetry direction (see Band Structure of Figure 7a, 10a). Conversely, Bands γ and δ display barrel-like shapes with prominent and open regions along the same M-L high symmetry direction, as illustrated in Figure 12.

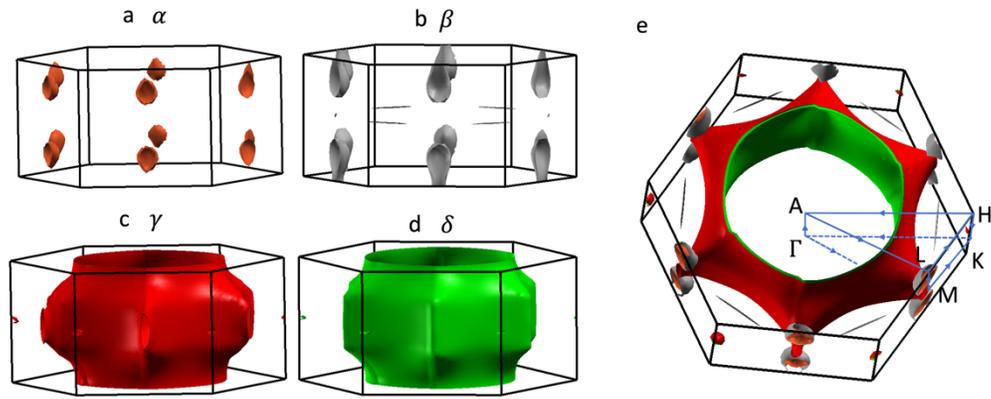

**Figure 12**. Projection of the Fermi surfaces of ferromagnetic GdV$_6$Sn$_6$ at the Fermi energy: (a) Band-$\alpha$, (b) Band-$\beta$, (c) Band-$\gamma$, (d) Band-$\delta$, and (e) all Fermi surface pockets. In (e) is depicted the high symmetry path followed to plot the band structure. All bands are depicted in different colors.